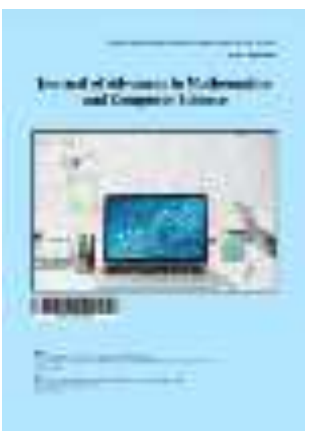

---

# Revisiting Linear Width: Rethinking the Relationship between Single Ideal and Linear Obstacle

**Takaaki Fujita [a*]**

[a] *Graduate School of Science and Technology, Gunma University, 1-5-1 Tenjin-Cho, Kiryu, Gunma 376-8515, Japan.*

***Author's contribution***

*The sole author designed, analyzed, interpreted and prepared the manuscript.*

---

## Abstract

The study of graph width parameters holds significant importance in the fields of graph theory and combinatorics. Among these parameters, linear-width stands out as a well-established and esteemed measure. The notions of single Ideal and Linear obstacle act as obstacles to achieving optimal linear-width in a connectivity system. In this succinct paper, we offer an alternative proof establishing the equivalence between single ideal and linear obstacle.

*Keywords: Linear width; single ideal; linear obstacle; connectivity system.*

---

**Corresponding author: Email: t171d603@gunma-u.ac.jp;*

# 1 Introduction

The exploration of width parameters holds significant importance in the realms of graph theory and combinatorics, as evidenced by the plethora of publications dedicated to this subject (e.g., [1, 2-5, 6-11, 12-26]). Among these parameters, branch-width, a well-studied concept, has received considerable attention in numerous papers. Likewise, linear width, a constrained version of branch-width, has been thoroughly investigated in a multitude of publications. Therefore, the study of both branch-width and linear width assumes critical significance [27,28].

The concept of single Ideal, introduced in reference [29], serves as a modeling tool for fundamental mathematical "ideal" in Boolean algebra and topology. In the context of a connectivity system, single Ideal represents the dual concept of linear width (also see reference [30,31]). Additionally, the notion of linear obstacle on a connectivity system corresponds to the dual concept of linear width [3,21].

Building upon these findings, it is established that single Ideal and linear obstacle are equivalent. However, in this concise paper, we present an alternative proof of their equivalence. While the level of novelty may be modest, our objective is to make a valuable contribution to the advancement of research in areas such as Linear width.

# 2 Definitions in this Paper

This section provides mathematical definitions for each concept.

## 2.1 Symmetric submodular function and connectivity system

The definition of a symmetric submodular function is provided below. However, it is important to note that although symmetric submodular functions can generally take real values, this paper specifically focuses on the subset of functions that take only natural numbers.

**Definition 1:** Let $X$ be a finite set. A function $f: X \rightarrow \mathbb{N}$ is called symmetric submodular if it satisfies the following conditions:

$$\forall A \subseteq X, f(A) = f(X \backslash A).$$

$$\forall A, B \subseteq X, f(A) + f(B) \geq f(A \cap B) + f(A \cup B).$$

A symmetric submodular function possesses the following properties. This lemma will be utilized in the proofs of lemmas and theorems presented in this paper.

**Lemma 1 [10]:** A symmetric submodular function f satisfies:

1. $\forall A \subseteq X, f(A) \geq f(\emptyset) = f(X)$,
2. $\forall A, B \subseteq X, f(A) + f(B) \geq f(A \backslash B) + f(B \backslash A)$.

In this brief paper, a connectivity system is defined as a pair $(X, f)$ consisting of a finite set (an underlying set) X and a symmetric submodular function $f$. Throughout this paper, we use the notation $f$ to refer to a symmetric submodular function, a finite set (an underlying set) $X$, and natural numbers $k, m$. A set $A$ is said to be $k$-efficient if $f(A) \leq k$.

## 2.2 Single ideal on a connectivity system *(X,f)*

The definition of a single ideal on a connectivity system $(X,f)$ is given below.

**Definition 2 [2]:** Let *X* represent a finite set and *f* denote a symmetric submodular function. In a connectivity system *(X,f)*, the set family $S \subseteq 2^X$ is called a single ideal of order *k+1* if the following axioms hold true: (IB) For every $A \in S, f(A) \leq k$.

(IH) If $A, B \subseteq X$, A is a proper subset of *B*, and *B* belongs to *S*, then *A* belongs to *S*.

(SIS) If *A* belongs to *S*, $e \in X$, $f(\{e\}) \leq k$, and $f(A \cup \{e\}) \leq k$, then $A \cup \{e\}$ belongs to *S*.

(IW) *X* does not belong to *S*.

In this short paper, we also consider the following additional axiom:

(IE) For each *k*-efficient subset *A* of *X*, exactly one of *A* or $(X \backslash A)$ is in *S*.

It has been shown in literature [2] that the linear width of *(X,f)* is at least *k+1* if and only if there exists a single ideal on *(X,f)* of order *k+1* that satisfies axiom (S4).

### 2.3 Linear obstacle on a connectivity system *(X,f)*

The definition of Linear obstacle is shown below. This concept is deep relation to *(k,m)*-obstacle in literature [6].
**Definition 3 [26]:** Let *X* represent a finite set and *f* denote a symmetric submodular function. In a connectivity system *(X,f)*, the set family $O \subseteq 2^X$ is called a linear obstacle of order *k + 1* if the following axioms hold true:

(O1) $A \in O, f(A) \leq k,$

(O2) $A \subseteq B \subseteq X, B \in O, f(A) \leq k \Rightarrow A \in O,$

(O3) $A, B, C \subseteq X, A \cup B \cup C = X, A \cap B = \emptyset, f(A) \leq k, f(B) \leq k, |C| \leq 1 \Rightarrow$ either $A \in O$ or $B \in O$.

## 3 Results: Equivalence between Single Ideal and Linear Obstacle

The result of this short paper is below.

**Theorem 1.** Let *X* represent a finite set and *f* denote a symmetric submodular function. Assuming that $f(\{e\}) \leq k$ for every $e \in X$, *S* is a single ideal of order *k+1* on *(X,f)* satisfying the additional axiom (IE) if and only if *S* is a linear obstacle of *order k+1 on (X,f)*.

**Proof of Theorem 1:**

(⇒) If *S* is a single ideal of order *k + 1* satisfying the additional axiom (IE), then *S* is a linear obstacle of order *k + 1*.

Axiom (O1) is clearly true. Axiom (O2) follows from axiom (IE).

To show axiom (O3), it is clear from axiom (IE) when $/C/ = 0$. When $/C/ = 1$, it is obvious from axiom (IE) if either $C \subseteq A$ or $C \subseteq B$. Therefore, consider the case where both $C \not\subseteq A$ and $C \not\subseteq B$ hold. Assume, without loss of generality, that $A \notin S$ and $B \notin S$, or $A \in S$ and $B \in S$. Here, we use the fact that either $A \notin S$ or $A \notin S$ holds, following from axiom (IE).

When $A \notin S$ and $B \notin S$, we have $(X \backslash A) = B \cup C \in S$. Since $f(B) \leq k$ and $B \subseteq B \cup C$, axiom (IH) implies $B \in S$, leading to a contradiction. When A $\in$ S and B $\in$ S, we have $(X \backslash A) = B \cup C \notin$ S. On the other hand, from axiom (SIS), we have $B \cup C \in S$, which leads to a contradiction.

(⇐) If $S$ is a linear obstacle of order $k + 1$, then $S$ is a single ideal of order $k + 1$ satisfying the additional axiom (IE).

Axiom (IH) and (IB) is obvious.

To show axiom (IE), assume $f(A) \leq k$. Since $A \cup (X \backslash A) = X$ and $A \cap (X \backslash A) = \emptyset$, either $A \in S$ or $(X \backslash A) \in S$ follows from axiom (O3).

To show axiom (SIS), assume $A \in S$ and $f(A \cup \{e\}) \leq k$. Then, we have $f((X \backslash A) \cap (X \backslash \{e\})) = f(A \cup \{e\}) \leq k$, implying $f((X \backslash A) \cap (X \backslash \{e\})) \leq k$. Since $A \in S$ and axiom (O1) hold, we have $f(A) \leq k$. From $A \cup ((X \backslash A) \cap (X \backslash \{e\})) \cup \{e\} = X$ and $A \cap ((X \backslash A) \cap (X \backslash \{e\})) = \emptyset$, either $A \in S$ or $(X \backslash A) \cap (X \backslash \{e\}) \in S$ follows from axiom (O3). Since A ∈ S, we obtain $A \cap (X \backslash \{e\}) \notin S$. Using the previously shown axiom (IE), we obtain $A \cup \{e\} \in S$.
To show axiom (IW), assume $X \in S$, which leads to a contradiction. Using Lemma 1, we obtain $f(X) = f(\emptyset) \leq k$. Using the previously shown axiom (IE) with $A = X$ and $B = \emptyset$, either $X \in S$ or $\emptyset \in S$ follows. Since $X \in S$, we obtain $\emptyset \notin S$, contradicting axiom (IH) which implies $\emptyset \in S$. This proof is completed.

## Competing Interests

Author has declared that no competing interests exist.

---